\documentclass[twocolumn]{aastex631}
\usepackage{amsmath}

\usepackage{longtable}

\usepackage[T1]{fontenc} 
\usepackage[utf8]{inputenc} 
\usepackage{amssymb}

\newcommand{\Princeton}{Department of Astrophysical Sciences, Princeton University, 4 Ivy Lane, Princeton, NJ 08540, USA}
\newcommand{\SAGAN}{NASA Sagan Fellow}

\received{June 1, 2019}
\revised{January 10, 2019}
\accepted{\today}
\submitjournal{\apjl}

\shorttitle{The Aligned Orbit of TOI-3362b}
\shortauthors{Espinoza-Retamal et al.}

\begin{document}

\title{The Aligned Orbit of the Eccentric Proto Hot Jupiter TOI-3362b\footnote{Based on observations made with ESO Telescopes at the La Silla Paranal Observatory under programme ID 110.23Y8.004}}

\correspondingauthor{Juan I. Espinoza-Retamal}
\email{jiespinozar@uc.cl}

\author[0000-0001-9480-8526]{Juan I.\ Espinoza-Retamal}
\affiliation{Instituto de Astrofísica, Pontificia Universidad Católica de Chile, Av. Vicuña Mackenna 4860, 782-0436 Macul, Santiago, Chile}

\author[0000-0002-9158-7315]{Rafael Brahm}
\affiliation{Facultad de Ingeniería y Ciencias, Universidad Adolfo Ibáñez, Av. Diagonal las Torres 2640, Peñalolén, Santiago, Chile}
\affiliation{Millennium Institute for Astrophysics, Chile}
\affiliation{Data Observatory Foundation, Chile}

\author[0000-0003-0412-9314]{Cristobal Petrovich}
\affiliation{Instituto de Astrofísica, Pontificia Universidad Católica de Chile, Av. Vicuña Mackenna 4860, 782-0436 Macul, Santiago, Chile}
\affiliation{Millennium Institute for Astrophysics, Chile}

\author[0000-0002-5389-3944]{Andr\'es Jord\'an}
\affiliation{Facultad de Ingeniería y Ciencias, Universidad Adolfo Ibáñez, Av. Diagonal las Torres 2640, Peñalolén, Santiago, Chile}
\affiliation{Millennium Institute for Astrophysics, Chile}
\affiliation{Data Observatory Foundation, Chile}
\affil{El Sauce Observatory -- Obstech, Chile}

\author[0000-0001-7409-5688]{Guðmundur Stefánsson} 
\affil{\SAGAN}
\affil{\Princeton}

\author[0000-0002-7444-5315]{Elyar Sedaghati} 
\affiliation{European Southern Observatory (ESO), Av. Alonso de Córdova 3107, 763 0355 Vitacura, Santiago, Chile}
\affiliation{Facultad de Ingeniería y Ciencias, Universidad Adolfo Ibáñez, Av. Diagonal las Torres 2640, Peñalolén, Santiago, Chile}
\affiliation{Millennium Institute for Astrophysics, Chile}

\author[0000-0002-5945-7975]{Melissa J. Hobson} 
\affiliation{Millennium Institute for Astrophysics, Chile}
\affiliation{Max-Planck-Institut für Astronomie, Königstuhl 17, D-69117 Heidelberg, Germany}

\author[0000-0003-2186-234X]{Diego J. Muñoz} 
\affiliation{Facultad de Ingeniería y Ciencias, Universidad Adolfo Ibáñez, Av. Diagonal las Torres 2640, Peñalolén, Santiago, Chile}
\affiliation{Millennium Institute for Astrophysics, Chile}
\affiliation{Data Observatory Foundation, Chile}
\affil{Center for Interdisciplinary Exploration and Research in Astrophysics (CIERA) and Department of Physics and Astronomy, Northwestern University, 2145 Sheridan Road, Evanston, IL 60208, USA}
\affiliation{Department of Astronomy and Planetary Science, Northern Arizona University, Flagstaff, AZ 86011, USA}

\author[0009-0009-2966-7507]{Gavin Boyle} 
\affil{El Sauce Observatory -- Obstech, Chile}
\affil{Cavendish Laboratory, J J Thomson Avenue, Cambridge, CB3 0HE, UK}

\author[0000-0002-6477-1360]{Rodrigo Leiva} 
\affiliation{Millennium Institute for Astrophysics, Chile}
\affiliation{Instituto de astrofísica de Andalucía, CSIC, Glorieta de la Astronomía s/n, 18008 Granada, Spain}

\author[0000-0001-7070-3842]{Vincent Suc} 
\affiliation{Millennium Institute for Astrophysics, Chile}
\affiliation{Facultad de Ingeniería y Ciencias, Universidad Adolfo Ibáñez, Av. Diagonal las Torres 2640, Peñalolén, Santiago, Chile}
\affil{El Sauce Observatory -- Obstech, Chile}

\begin{abstract}
High-eccentricity tidal migration predicts the existence of highly eccentric proto-hot Jupiters on the ``tidal circularization track,'' meaning that they might eventually become hot Jupiters, but that their migratory journey remains incomplete. Having experienced moderate amounts of tidal evolution of their orbital elements, proto-hot Jupiters systems can be powerful test beds for the underlying mechanisms of eccentricity growth. Notably, they may be used for discriminating between variants of high-eccentricity migration, each predicting a distinct evolution of misalignment between the star and the planet's orbit. We constrain the spin-orbit misalignment of the proto-hot Jupiter TOI-3362b with high-precision radial velocity observations using ESPRESSO at VLT. The observations reveal a sky-projected obliquity $\lambda=1.2_{-2.7}^{+2.8}$ deg and constrain the orbital eccentricity to $e=0.720\pm0.016$, making it one of the most eccentric gas giants for which the obliquity has been measured. Although the large eccentricity and the striking orbit alignment of the planet are puzzling, we suggest that ongoing coplanar high-eccentricity migration driven by a distant companion is a possible explanation for the system's architecture. This distant companion would need to reside beyond 5 au at 95\% confidence to be compatible with the available radial velocity observations.
\end{abstract}


\section{Introduction}
\label{sec:intro}

Ever since the discovery of 51 Pegasi b \citep{Mayor95}, the existence of hot Jupiters (HJs) has continued to elude a definitive explanation (see \citet{Dawson18} for a review on the possible origins of HJs). Generally, the theoretical channels proposed to explain these close-in planets fall into two categories. One is high-eccentricity (high-$e$) tidal migration, in which proto-HJs are first launched into highly eccentric orbits --- either through planet-planet scattering \cite[e.g.,][]{Rasio96,beauge2012}, von Zeipel-Lidov–Kozai oscillations \citep[e.g.,][]{Wu03,Fabrycky07,Naoz11}, or secular interactions \citep[e.g.,][]{Wu11,Petrovich15}--- after which tidal friction acts to circularize and shrink the orbits. The other proposed channel posits that HJs acquire their current orbits while still embedded in their parental gas disks, either by in situ formation \citep[e.g.,][]{Batygin16,Boley16} or by formation beyond the ice line, followed by inward migration driven by nebular tides \citep[e.g.,][]{Goldreich80,Lin86}. 

If close-in giant planets are indeed the result of high-$e$ tidal migration, we would expect to catch at least some planets on the tidal circularization track \citep{socrates2012}. Arguably the clearest example of such a system is HD80606b \citep{Naef01}, which has a distant stellar companion and exhibits extreme orbital eccentricity $e=0.93$, and high stellar obliquity \citep{Moutou09, Pont09, Winn09, Hebrard10}, and is thus naturally explained by the von Zeipel-Lidov-Kozai mechanism \citep{Wu03}. 
 
Similarly, TOI-3362b,  with a reported orbital period $P=18.1$ days and eccentricity $e=0.815$ \citep{Dong21}, may also be considered to be part of the small group of gas giants known to be undergoing high-$e$ tidal migration. Nonetheless, due to the absence of detected (stellar or planetary) companions, coupled with the unresolved matter of stellar obliquity, conclusively identification of the mechanisms that have driven TOI-3362b toward high eccentricity has not been possible to date.
 
In this Letter, we present further orbital characterization of the TOI-3362b system, obtained from measurements of the Rossiter–McLaughlin (RM) effect and precise long-term RV follow-up observations. Precise in-transit spectroscopic observations with ESPRESSO revealed that the planet's orbital axis is aligned with the stellar rotation axis with better than three-degree precision and the long-term RV follow-ups obtained with FEROS shows no clear features of long-period companions in the system. 

\section{Observations}\label{sec:obs}

\subsection{Transit Spectroscopy}

We observed a single transit of TOI-3362b with the ESPRESSO spectrograph \citep{Pepe20} on the night of the 27th of December 2022, between 04:31 and 08:16 UT. ESPRESSO is the highly-stabilized, fiber-fed cross-dispersed echelle spectrograph installed at the Incoherent Combined Coudé Focus of ESO's Paranal Observatory in Chile. It covers the wavelength range of 380 to 788 nm at a resolving power of $R\approx$ 140,000 in single-UT High-Resolution mode. We obtained 38 spectra of the host star during the primary transit with UT2 at an exposure time of 310 s. The observations were performed under clear sky conditions, with atmospheric seeing in the range of 0.58 $-$ 1.95$^{\prime\prime}$. The spectra have a median S/N of 54 at 550 nm and a median RV uncertainty of 12.4 m/s. The data were reduced with the dedicated data reduction pipeline (v. 2.4.0), including all standard reduction steps, which also provides RVs by fitting a Gaussian model to the calculated cross-correlation function (CCF). The CCF is calculated at steps of 0.5 km/s (representing the sampling of the spectrograph) for $\pm 100$ km/s centered on the estimated systemic velocity. The RM component of the RVs is presented in Figure \ref{fig:RM}.

\subsection{Photometry}\label{subsec:photometry}

\subsubsection{Observatoire Moana}

Simultaneously with ESPRESSO observations, we observed the transit using the station of the Observatoire Moana located in El Sauce (ES) Observatory in Chile. This station consists of a 0.6 m CDK robotic telescope coupled to an Andor iKon-L 936 deep depletion 2k $\times$ 2k CCD with a scale of 0.67$^{\prime\prime}$ per pixel. We used a Sloan $r'$ filter, and the exposure time was 14 s. The drift of the stars in the CCD during the sequence was smaller than 5 pixels, and the airmass ranged from 1.8 to 1.2. We processed the data with a dedicated pipeline that automatically performs the CCD reduction steps, followed by the measurement of the aperture photometry for the brightest stars in the field. The pipeline also generates the differential light curve of the target star by identifying the optimal comparison stars based on color, brightness, and proximity to the target. The light curve data is displayed in Figure \ref{fig:TR} along with the optimal model.

\subsubsection{TESS}

TOI-3362 was observed by the Transiting Exoplanet Survey Satellite \citep[TESS,][]{Ricker2015} in Sectors 9 and 10 (Year 1) at a cadence of 30 minutes, in Sectors 36 and 37 (Year 3) at a cadence of 10 minutes, and in Sectors 63 and 64 (Year 5) at a cadence of 2 minutes. We searched and downloaded all the light curves using \texttt{lightkurve} \citep{Lightkurve}. TESS light curves were processed by the Science Processing Operations Center (SPOC) pipeline \citep{Jenkins16}. The Presearch Data Conditioning (PDC) component of the SPOC pipeline corrects the light curves for pointing or focus-related instrumental signatures, discontinuities resulting from radiation events in the CCD detectors, outliers, and flux contamination. TESS data is displayed in Figure \ref{fig:TR} along with the best-fit transit model.

\subsubsection{Archival Observations}

For this work, we also made use of the ground-based light curves published in \citet{Dong21}, who presented  4 photometric transit observations. Two of them were taken with the Antarctica Search for Transiting ExoPlanets (ASTEP) program \citep{Guillot15,Mekarnia16}, which observed the transits of TOI-3362b on the nights of 10th August 2020 and 28th August 2020. This 0.4 m telescope is equipped with an FLI Proline science camera with a KAF-16801E, $4096\times4096$ front-illuminated CCD with a scale of 0.93" per pixel, and it is located on the east Antarctic plateau. Observations were performed in a band that is similar to the $R_c$ in transmission. Also, they observed one transit of TOI-3362b with the 1.0 m telescope at the Las Cumbres Observatory Global Telescope \citep[LCOGT,][]{Brown13} Siding Spring Observatory (SSO) node in New South Wales, Australia. This telescope is equipped with a $4096\times4096$ Sinistro camera with a scale of 0.389" per pixel. The transit was observed with the Pan-STARSS $z$-short and Bessell $B$ filters on the night of 7th February 2021.

\subsection{Spectroscopy}

\subsubsection{FEROS}

We obtained 7 spectra of TOI-3362 with FEROS \citep{Kaufer99} between 23rd November 2021 and 6th June 2023. FEROS is an echelle spectrograph mounted at the MPG/ESO 2.2 m telescope located at La Silla Observatory, Chile. These observations were performed to identify long-period companions to TOI-3362b. The exposure times were set to 300 s, yielding an average S/N per spectral resolution element of 65. FEROS spectra were reduced, extracted, and analyzed with the \texttt{ceres} pipeline \citep{Brahm2017}. The obtained FEROS RVs are displayed in Figure \ref{fig:RV}. No signatures of additional planets are identified from these observations.

\subsubsection{Archival Observations}

For this work, we also made use of the radial velocities published in \citet{Dong21}. The first dataset was obtained from the analysis of 21 spectra taken with the CHIRON echelle spectrograph \citep{Tokovinin13} mounted on the SMARTS 1.5 m telescope at Cerro Tololo in Chile. These observations were made between 15th April 2021 and 12th May 2021. The second dataset was obtained from the analysis of 9 spectra taken with the Minerva-Australis array \citep{Addison19}. Minerva-Australis is an array of four identical 0.7 m telescopes linked via fiber feeds to a single KiwiSpec echelle spectrograph on Mt. Kent Observatory, Australia. These observations were made between 16th May 2021 and 30th May 2021.

\section{Stellar Parameters}\label{sec:star}
We obtained the stellar parameters by following the procedure presented in \citet{brahm:2019}. We use a two-step iterative process. The first process consists of obtaining the atmospheric parameters ($T_{eff}$, $\log{g}$, [Fe/H], and $v \sin{i}$) of the host star from a high-resolution spectrum using the \texttt{zaspe} code \citep{zaspe}. This code compares the observed spectrum to a grid of synthetic spectra and determines reliable uncertainties that take into account systematic mismatches between the observations and the imperfect theoretical models. For this analysis, we used the co-added out-of-transit ESPRESSO spectra to obtain the atmospheric parameters. The second step consists of obtaining the physical parameters of the star by using publicly available broad-band magnitudes of the star and comparing them with those produced by different PARSEC stellar evolutionary models \citep{parsec} by taking into account the distance to the star computed from Gaia DR2 \citep{dr2} parallax. This procedure delivers a new value of $\log{g}$ that is held fixed in a new run of \texttt{zaspe}. We iterate between the two procedures until reaching convergence in the $\log{g}$. The obtained parameters are presented in Table \ref{stpars}.

\begin{deluxetable}{lrr}
\tablecaption{Stellar properties$^a$ of TOI-3362.  \label{stpars}}
\tablecolumns{3}
\tablewidth{0pt}
\tablehead{Parameter & Value & Reference}
\startdata
RA \dotfill (J2015.5) &  10h23m56.19s & Gaia DR2\\
Dec \dotfill (J2015.5) & -56d50m35.22s & Gaia DR2\\
pm$^{\rm RA}$ \hfill (mas/yr) & -24.12 $\pm$ 0.05 & Gaia DR2\\
pm$^{\rm DEC}$ \dotfill (mas/yr) & 7.72 $\pm$ 0.05 & Gaia DR2\\
$\pi$ \dotfill (mas)& 2.79 $\pm$ 0.01 & Gaia DR2 \\ 
\hline
T \dotfill (mag) & 10.382 $\pm$ 0.006 & TICv8\\
B  \dotfill (mag) & 11.643 $\pm$ 0.12 &  APASS$^b$\\
V  \dotfill (mag) & 10.86 $\pm$ 0.010  & APASS\\
G  \dotfill (mag) & 10.7051 $\pm$ 0.002 &  Gaia DR2$^c$\\
G$_{BP}$  \dotfill (mag) & 10.948 $\pm$ 0.005 & Gaia DR2\\
G$_{RP}$  \dotfill (mag) & 10.335 $\pm$ 0.003 & Gaia DR2\\
J  \dotfill (mag) &  9.94 $\pm$ 0.02 & 2MASS$^d$\\
H  \dotfill (mag) &  9.72 $\pm$ 0.02 &  2MASS\\
K$_s$  \dotfill (mag) & 9.69 $\pm$ 0.02 & 2MASS\\
\hline
$T_{eff}$  \dotfill (K) & 6800 $\pm$ 100  & This work\\
$\log{g}$ \dotfill (dex) & 4.14 $\pm$ 0.01 & This work\\
$[Fe/H]$ \dotfill (dex) & 0.24 $\pm$ 0.05  & This work\\
$v\sin{i}$ \dotfill (km/s) & 21.9 $\pm$ 0.8 & This work\\
$M_{\star}$ \dotfill ($M_{\odot}$) & 1.53 $\pm$ 0.02 & This work\\
$R_{\star}$ \dotfill ($R_{\odot}$) & 1.75 $\pm$ 0.02 & This work\\
L$_{\star}$ \dotfill (L$_{\odot}$) & 5.9 $\pm$ 0.3 & This work\\
A$_{V}$ \dotfill (mag) & 0.26 $\pm$ 0.06 & This work\\
Age \dotfill (Gyr) & 1.3 $\pm$ 0.2 & This work\\
$\rho_\star$ \dotfill (g/cm$^{3}$) & 0.40 $\pm$ 0.02 & This work\\
\enddata
\tablecomments{$^a$ The stellar parameters computed in this work do not consider possible systematic differences among different stellar evolutionary models \citep{tayar:2022} and have underestimated uncertainties, $^b$\citet{apass},$^c$\citet{dr2},$^d$\citet{2mass}}.
\end{deluxetable}

\section{Photometric Analysis}\label{sec:photo}

To update the orbital ephemeris of TOI-3362b and look for Transit Timing Variations (TTVs), we modeled all the photometric data presented in Section \ref{subsec:photometry} using the \texttt{juliet} code \citep{Espinoza19}. \texttt{juliet} uses \texttt{batman} \citep{Kreidberg15} for the transit model and the \texttt{dynesty} dynamic nested sampler \citep{Speagle20} to perform bayesian analysis and explore the likelihood space to obtain posterior probability distributions. We placed uninformative priors on the transit parameters $R_p/R_{\star}$ and $b$, with an informative prior on the stellar density, that was constrained in Section \ref{sec:star}. We sampled the limb darkening parameters using the quadratic $q_1$ and $q_2$ from \citet{Kipping13} with uniform priors. We placed Gaussian priors for each transit mid-point based on the expected values calculated from the orbital period and $t_0$ from \citet{Dong21}, with uncertainties of 0.1 days. To account for variability in TESS light curves, we included a Matern-3/2 Gaussian Process (GP) as implemented in \texttt{celerite} \citep{Foreman-Mackey17} and available in \texttt{juliet}. Year 1, 3, and 5 had their own GP kernels to account for differences in variability captured in different epochs and cadences. From this analysis, we ruled out the presence of TTVs greater than $\sim5$ minutes and we obtained improved orbital ephemeris of TOI-3362b.

\begin{deluxetable*}{llcc}
\tablecaption{Summary of priors and resulting posteriors of the joint fit. \label{tab:fit}}
\tablewidth{70pt}
\tabletypesize{\scriptsize}
\tablehead{Parameter & Description & Prior & Posterior}
\startdata 
$\lambda$ & Sky-projected obliquity (deg) & $U(-180,180)$ & $1.2_{-2.7}^{+2.8}$\\
$v\sin{i}$ & Projected rotational velocity (km/s) & $U(0,30)$ & $20.2_{-1.6}^{+1.7}$\\
$P$ & Orbital period (days) & $N(18.095368,0.000012)$ & $18.095367_{-0.000008}^{+0.000008}$\\
$t_0$ & Transit midpoint (BJD) & $N(2458529.3277,0.0007)$ & $2458529.3279_{-0.0006}^{+0.0006}$\\
$\rho_\star$ & Stellar density (g/cm$^3$)& $N(0.40,0.02)$ & $0.40_{-0.02}^{+0.02}$\\
$b$ & Impact parameter & $U(0,1)$ & $0.57_{-0.05}^{+0.04}$\\
$R_p/R_\star$ & Radius ratio & $U(0,1)$ & $0.070_{-0.001}^{+0.001}$\\
$e$ & Eccentricity & $U(0,0.95)$ & $0.720_{-0.016}^{+0.016}$\\
$\omega$ & Argument of periastron (deg) & $U(0,360)$ & $60.6_{-7.1}^{+8.0}$\\
$K$ & RV semiamplitude (m/s) & $U(0,1000)$ & $338_{-27}^{+27}$\\
$a/R_\star$ & Scaled semimajor axis & - & $19.1_{-0.3}^{+0.3}$\\
$i$ & Orbital inclination (deg) & - & $84.25_{-0.31}^{+0.34}$\\
$a$ & Semimajor axis (au) & - & $0.155_{-0.003}^{+0.003}$ \\
$R_p$ & Planet radius ($R_J$) & - & $1.2_{-0.02}^{+0.02}$ \\
$M_p$ & Planet mass ($M_J$) & - & $4.0_{-0.4}^{+0.4}$\\
$\rho_p$ & Planet mean density (g/cm$^{3}$) & - & $3.0_{-0.3}^{+0.3}$ \\
\hline
$q_1^{\rm ESPRESSO}$ & ESPRESSO linear limb darkening parameter & $U(0,1)$ & $0.42_{-0.17}^{+0.22}$\\
$q_2^{\rm ESPRESSO}$ & ESPRESSO quadratic limb darkening parameter & $U(0,1)$ & $0.58_{-0.34}^{+0.29}$\\
$\beta$ & Intrinsic stellar line width (km/s) & $N(7.8,2.0)$ & $7.6_{-1.9}^{+1.9}$\\
$\gamma_{\rm ESPRESSO}$ & ESPRESSO RV offset (m/s) & $U(7500,8000)$ & $7637_{-33}^{+35}$\\
$\gamma_{\rm CHIRON}$ & CHIRON RV offset (m/s) & $U(6200,6600)$ & $6443_{-35}^{+35}$\\
$\gamma_{\rm MINERVA}$ & Minerva-Australis RV offset (m/s) & $U(7400,7800)$ & $7604_{-56}^{+55}$\\
$\gamma_{\rm FEROS}$ & FEROS RV offset (m/s) & $U(7800,8200)$ & $8029_{-35}^{+35}$\\
$\sigma_{\rm ESPRESSO}$ & ESPRESSO RV jitter (m/s) & $LU(10^{-3},100)$ & $0.1_{-0.1}^{+1.0}$\\
$\sigma_{\rm CHIRON}$ & CHIRON RV jitter (m/s) & $LU(10^{-3},100)$ & $0.3_{-0.3}^{+15.6}$\\
$\sigma_{\rm MINERVA}$ & Minerva-Australis RV jitter (m/s) & $LU(10^{-3},100)$ & $0.4_{-0.3}^{+13.2}$\\
$\sigma_{\rm FEROS}$ & FEROS RV jitter (m/s) & $LU(10^{-3},100)$ & $0.2_{-0.2}^{+7.2}$\\
\hline
$q_1^{\rm TESS}$ & TESS linear limb darkening parameter & $U(0,1)$ & $0.08_{-0.05}^{+0.10}$\\
$q_2^{\rm TESS}$ & TESS quadratic limb darkening parameter & $U(0,1)$ & $0.38_{-0.27}^{+0.37}$\\
$q_1^{r'}$ & $r'$ linear limb darkening parameter & $U(0,1)$ & $0.73_{-0.26}^{+0.19}$\\
$q_2^{r'}$ & $r'$ quadratic limb darkening parameter & $U(0,1)$ & $0.72_{-0.33}^{+0.20}$\\
$q_1^{B}$ & $B$ linear limb darkening parameter & $U(0,1)$ & $0.05_{-0.04}^{+0.10}$\\
$q_2^{B}$ & $B$ quadratic limb darkening parameter & $U(0,1)$ & $0.37_{-0.27}^{+0.37}$\\
$q_1^{z_s}$ & $z_s$ linear limb darkening parameter & $U(0,1)$ & $0.12_{-0.08}^{+0.15}$\\
$q_2^{z_s}$ & $z_s$ quadratic limb darkening parameter & $U(0,1)$ & $0.30_{-0.22}^{+0.35}$\\
$q_1^{R_c}$ & $R_c$ linear limb darkening parameter & $U(0,1)$ & $0.08_{-0.06}^{+0.12}$\\
$q_2^{R_c}$ & $R_c$ quadratic limb darkening parameter & $U(0,1)$ & $0.32_{-0.23}^{+0.37}$\\
$\sigma_{\rm TESS}^{\rm Y1}$ & TESS Year 1 photometric jitter & $LU(10^{-6},50)$ & $0.00142_{-0.00012}^{+0.00014}$\\
$\sigma_{\rm TESS}^{\rm Y3}$ & TESS Year 3 photometric jitter & $LU(10^{-6},50)$ & $0.00089_{-0.00006}^{+0.00006}$\\
$\sigma_{\rm TESS}^{\rm Y5}$ & TESS Year 5 photometric jitter & $LU(10^{-6},50)$ & $0.00002_{-0.00001}^{+0.00010}$\\
$\sigma_{r'}$ & $r'$ photometric jitter & $LU(10^{-6},50)$ & $0.00351_{-0.00021}^{+0.00022}$\\
$\sigma_{B}$ & $B$ photometric jitter & $LU(10^{-6},50)$ & $0.00140_{-0.00010}^{+0.00011}$\\
$\sigma_{z_s}$ & $z_s$ photometric jitter & $LU(10^{-6},50)$ & $0.00063_{-0.00013}^{+0.00012}$\\
$\sigma_{R_c}$ & $R_c$ photometric jitter & $LU(10^{-6},50)$ & $0.00174_{-0.00007}^{+0.00007}$\\
\enddata
\tablecomments{$U(a,b)$ denotes a uniform prior with a start value $a$ and end value $b$. $N(m,\sigma)$ denotes a normal prior with mean $m$, and standard deviation $\sigma$. $LU(a,b)$ denotes a log-uniform prior with a start value $a$ and end value $b$.}
\end{deluxetable*}

\section{Joint fit}\label{sec:fit}

\begin{figure*}[t]
\centering
\includegraphics[width=15cm]{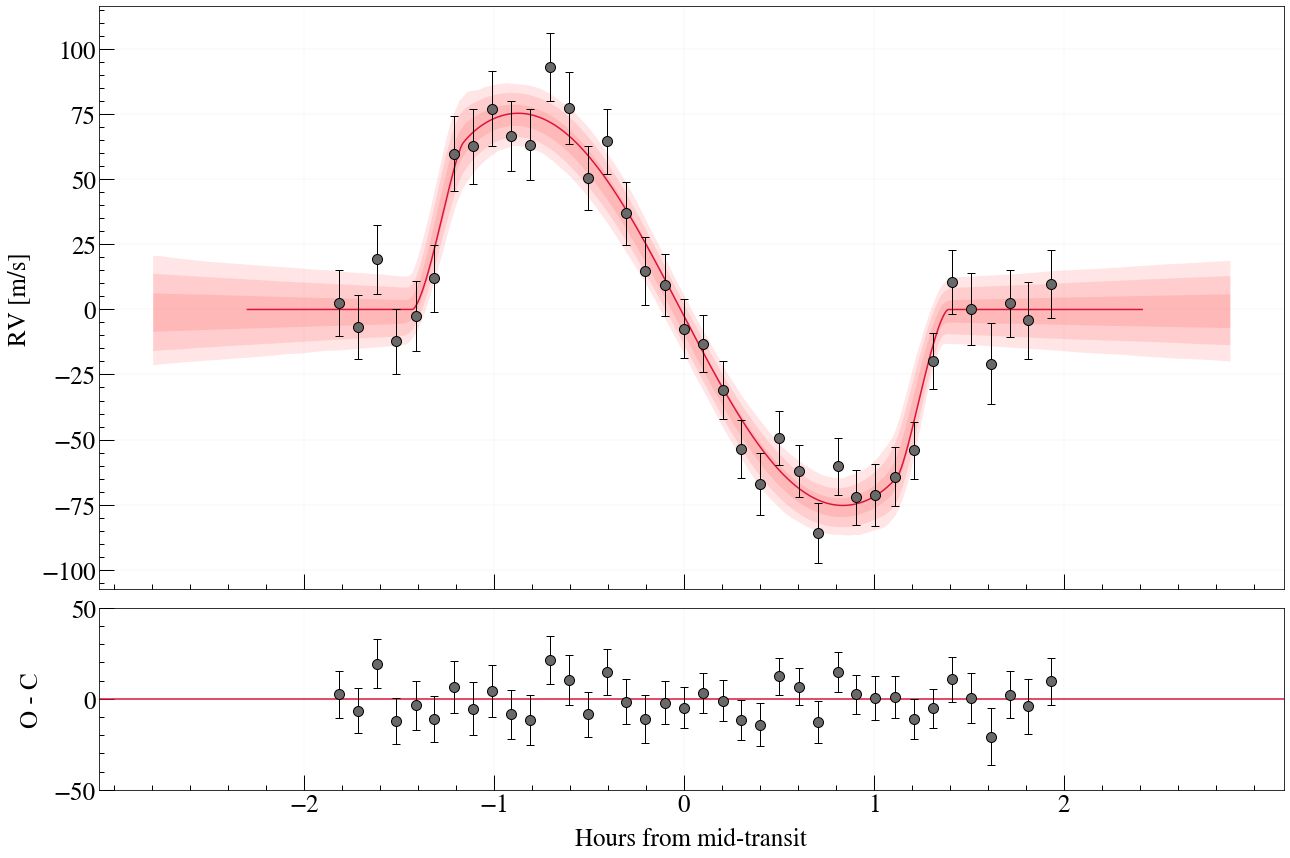}
    \caption{ESPRESSO observations of the RM effect of TOI-3362b with the Keplerian signal removed. The red line shows the best fit, and the red areas show 1, 2, and 3$\sigma$ models. The bottom panel shows the residuals. The ESPRESSO RVs of this figure without the Keplerian signal removed are available as the Data behind the Figure.}
    \label{fig:RM}
\end{figure*}

\begin{figure*}[b]
    \centering
    \includegraphics[width = 15cm]{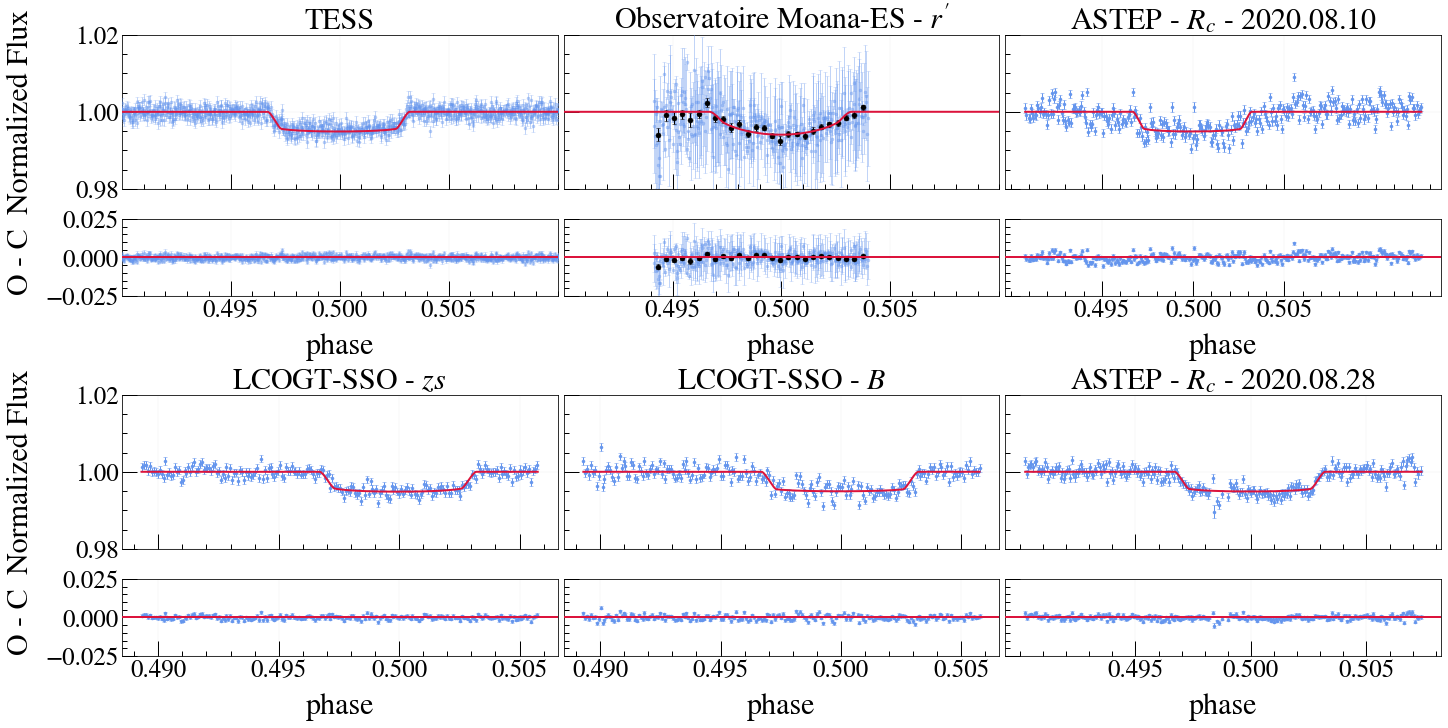}
    \caption{TESS and ground-based transit photometry of TOI-3362b overplotted with best-modeled lightcurves. Residuals are shown for all transits. TESS light curves of different years are shown all together. Observatoire Moana - ES light curve is also shown with 10min-binned data in black, and it is available as the Data behind the Figure.}
    \label{fig:TR}
\end{figure*}

\begin{figure*}[t]
    \centering
    \includegraphics[width =18cm]{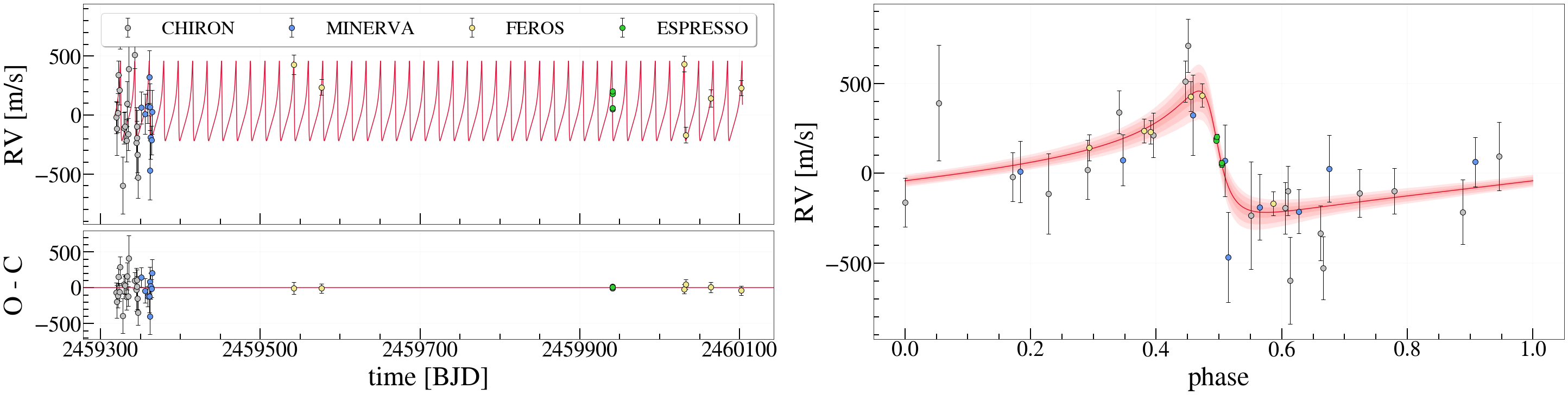}
    \caption{\textit{Left:} Out-of-transit radial velocities of TOI-3362 taken with different instruments. The red curve shows the best model. Residuals are shown at the bottom. \textit{Right:} Out-of-transit radial velocity curve in phase. The red curve shows the best model, and the red regions show 1, 2, and 3$\sigma$ models. FEROS data is available as the Data behind the Figure.}
    \label{fig:RV}
\end{figure*}

To precisely constrain the parameters of TOI-3362b and its orbit, we jointly modeled all observations presented in Section \ref{sec:obs}. In total, our model considered 38 parameters, 10 physical and 28 instrumental (see Table \ref{tab:fit}). We used \texttt{batman} \citep{Kreidberg15} to model all the light curves. To reduce the computational cost, we only considered the detrended TESS lightcurve in phases between 0.48 and 0.52 ($\sim$ 5 times the transit duration, see Figure \ref{fig:TR}). For the limb darkening, we considered a quadratic law using the parametrization from \citet{Kipping13}. To model the out-of-transit radial velocities and RM effect, we used the \texttt{rmfit} package \citep{Stefansson20,Stefansson22}, which uses the framework from \citet{Hirano10} to model the anomalous RVs seen during the RM effect and \texttt{radvel} \citep{Fulton18} to model the Keplerian out-of-transit RV curve. We removed one FEROS and two CHIRON measurements that were taken during the transits of the planet, and we did not use them during the analysis. To account for the instrumentation offsets and systematics, we included independent RV offsets and log-uniform jitter terms for each instrument.

We placed uninformative priors for almost all parameters, except for the orbital period and time of mid-transit, which had been constrained in Section \ref{sec:photo}, and for the stellar density that was constrained in Section \ref{sec:star}. Likewise, for $\beta$, which is the total line width accounting for both instrumental resolution and macroturbulence velocity. We considered an instrumental broadening of 2.15 km/s because of the ESPRESSO resolution, and following \citet{Albrecht12} we used the macroturbulence law for hot stars from \citet{Gray84}, which yields a broadening of $\sim7.5$ km/s for $T_{eff}\sim$ 6,800 K. We added those broadening values in quadrature to set our prior for $\beta$, with an uncertainty of 2 km/s. We also tested with an uninformative prior, which resulted in the same posterior values, demonstrating that the posteriors are insensitive to $\beta$. All priors and resulting posteriors are shown in Table \ref{tab:fit}. To estimate the bayesian posteriors and evidences of our model, we used the \texttt{dynesty} dynamic nested sampler. Since the number of parameters was 38, we considered 3,500 live points to ensure convergence. Figure \ref{fig:corner} shows the joint posterior distributions and histograms for the physical parameters of our model. All parameters appear well-behaved and approximately Gaussian in distribution.

We performed a series of tests to verify our results. First, we tried \texttt{emcee} \citep{Foreman-Mackey13} to sample the posteriors. Second, we performed a joint fit of the photometry and the out-of-transit RVs with \texttt{juliet}. Finally, we carried out a fully independent analysis using the \texttt{exoplanet} \citep{Foreman21} package implementing the RM effect using the formalism of \citet{Hirano11}. All of them returned a completely consistent set of orbital and stellar obliquity parameters.

Figures \ref{fig:RM}, \ref{fig:TR}, and \ref{fig:RV} show the different datasets together with the best-fit model. We found a sky-projected obliquity $\lambda=1.2_{-2.7}^{+2.8}$ deg, which means that the TOI-3362b is perfectly aligned with its host star. This result is surprising given the high eccentricity $e=0.720\pm0.016$ of its orbit. We also found lower values of eccentricity and inclination than the ones reported by \citet{Dong21} of $0.815_{-0.032}^{+0.023}$ and $89.140_{-0.668}^{+0.584}$ deg, respectively. That is mainly a result of the additional datasets included in this work. In particular, FEROS, ESPRESSO, and TESS Year 5 measurements allowed us to better constrain the orbit of the planet. But in general, the parameters are consistent with the previously reported ones. The baseline of the RV measurements used in this work is $\sim2$ years, and residuals show no clear evidence of companions (see Figure \ref{fig:RV}). Therefore, based on the residuals, we rule out the presence of a companion with at least the same mass in orbits with $a\lesssim4.5$ au (see Section \ref{subsec:companions}).

\begin{figure*}[t]
    \centering
    \includegraphics[width = 18cm]{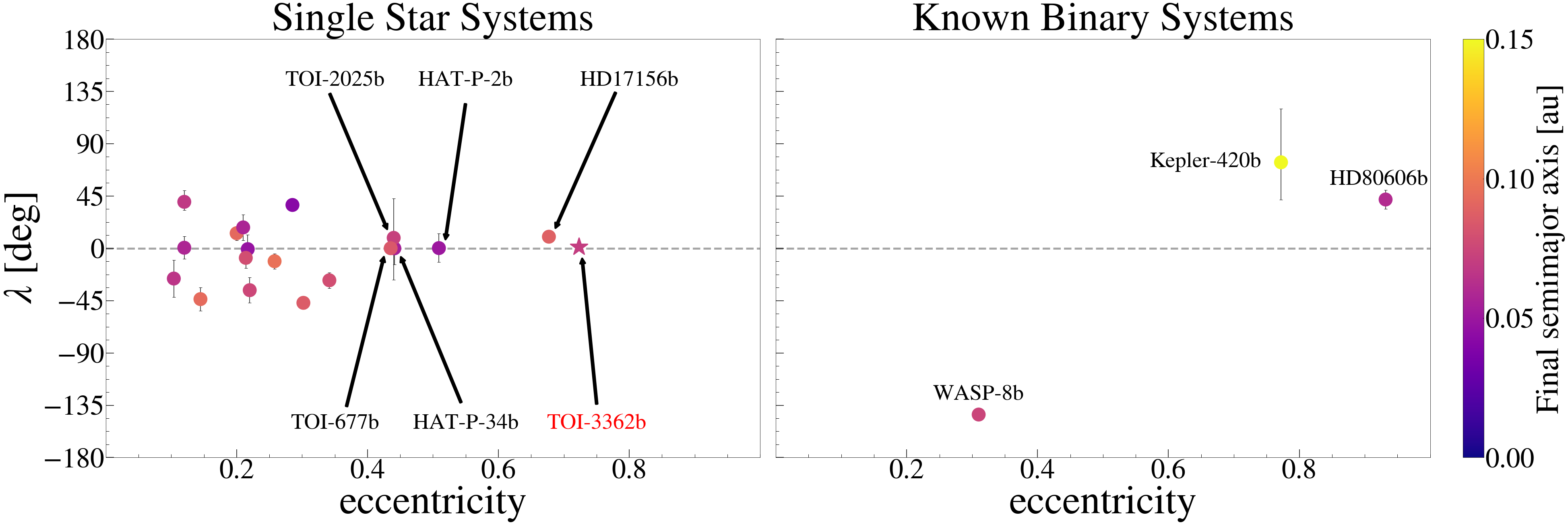}
    \caption{Sky-projected obliquity versus the eccentricity of all giant planets ($M_p>0.2\, M_J$ and $R_p>8\,R_{\oplus}$) with $e>0.1$ colored by their final semimajor axes $a_{\rm final} = a(1-e^2)$ from TEPCat \citep{Southworth11}. TOI-677b was also measured using ESPRESSO \citep{Sedaghati23}. The left panel shows single stars, while the right panel shows known binary systems. All planets, excepting Kepler-420b, are or will be hot Jupiters with enough time.}
    \label{fig:scatter}
\end{figure*}

\begin{figure}
    \centering
    \includegraphics[width = \columnwidth]{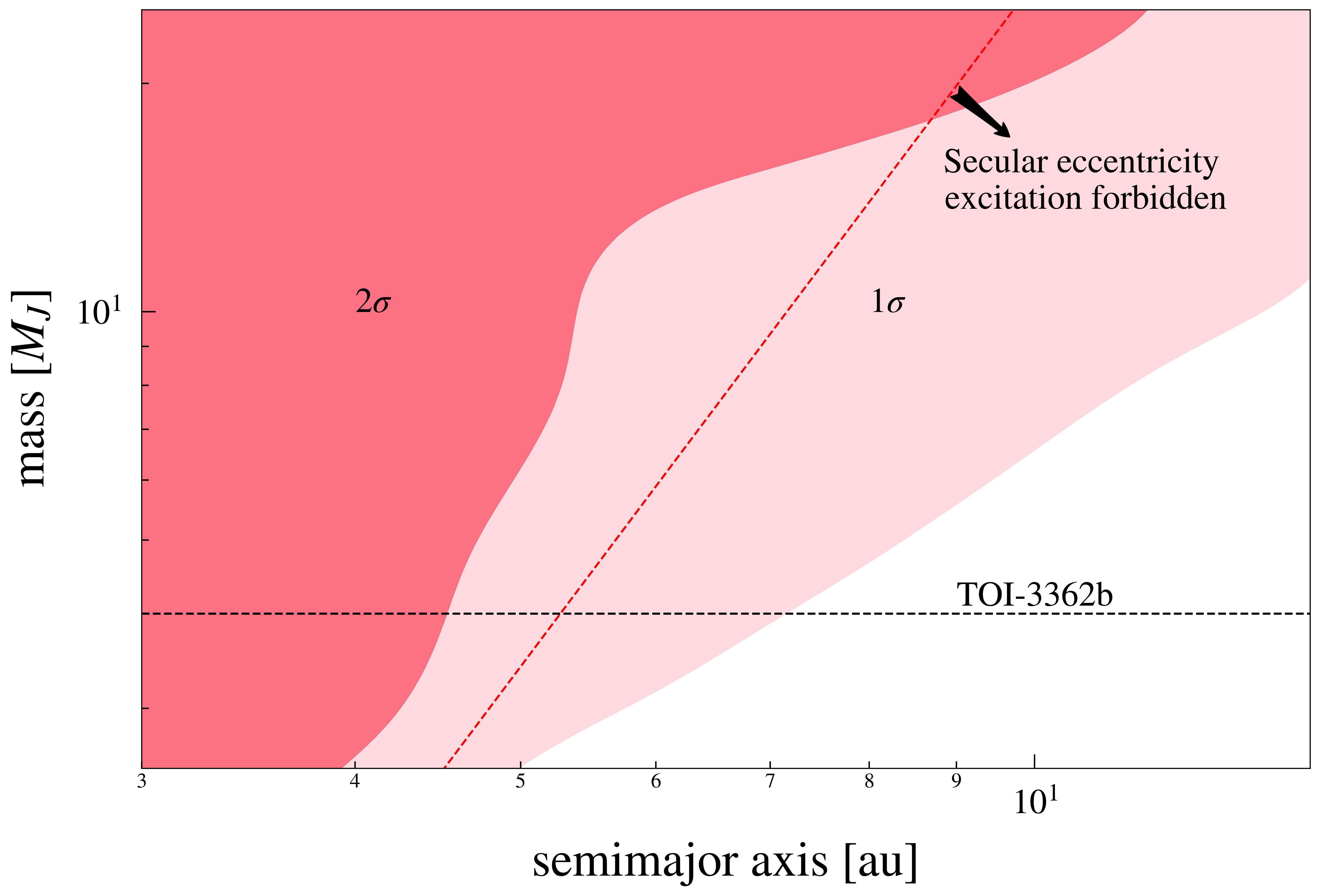}
    \caption{Mass versus semimajor axis diagram for the possible coplanar outer companions to TOI-3362b. Red regions indicate zones where we can rule out the companion at 1 and $2\sigma$ confidence levels. The black dashed line shows the mass of TOI-3362b. The line $\epsilon_{\rm gr}=1$ (Eq. \ref{eq:epsilon}) is shown in red dashed, demarking the regions where secular excitation is quenched by relativistic precession ($\epsilon_{\rm gr}>1$). Since the companion might lie in the region where secular eccentricity excitation is forbidden, we discard it as responsible for the current high eccentricity of TOI-3362b.} 
    \label{fig:companions}
\end{figure}

\section{Discussion}\label{sec:discussion}

The values of the semimajor axis and eccentricity obtained here are consistent with TOI-3362b being a proto HJ undergoing high-$e$ tidal migration. The approximate final orbital separation is  $a_{\rm final} = a(1-e^2) \approx 9.2\,R_{\star}$ or $0.07$ au, corresponding to an orbital period of $P_{\rm final}\approx6$ days. We estimate a tidal circularization timescale of the planet following \citet{Goldreich66}:
\begin{equation}
    \tau_{\rm circ}\equiv -\frac{e}{\dot{e}}=\frac{2P}{63\pi}Q'_p\frac{M_p}{M_\star} \left( \frac{a}{R_p}\right)^5 F(e),
\end{equation}
where $Q'_p$ is the planet's modified tidal quality factor \citep[e.g.,][]{Goldreich66,Ogilvie07} and $F(e)$ is an eccentricity-dependent correction factor \citep{Hut81}. For $e=0.720$, and assuming pseudo-synchronous rotation of the planet, we have $F(e)\approx0.004$. Further, assuming that $Q'_p= 10^5-10^6$, we have $\tau_{\rm circ}\sim 0.8-8$ Gyr for TOI-3362b. This timescale is longer than that reported by \citet{Dong21}, who inferred $\tau_{\rm circ}\sim2.7$ Gyr assuming an eccentricity of 0.815 and $Q'_p=10^6$. Given that the age of TOI-3362 is $1.3\pm0.2$ Gyr, it is not surprising that we observe the planet in an eccentric orbit. Moreover, the planetary orbit may not have enough time to fully circularize before the host star exits the main sequence. Nevertheless, it could undergo sufficient orbital decay to meet the criteria for classification as an eccentric HJ ($P<10$ days). Indeed, direct numerical integration\footnote{We integrate the eccentricity evolution at constant $a_{\rm final}$, from $e=0.72$ to $e=0.53$ at which point the orbit shrunk to 0.11 au or an orbital period of 10 days.} shows that the orbit can shrink down to 0.11 au in a timespan of $0.3-3$ Gyr.

In Figure \ref{fig:scatter}, we compare the orbital properties of TOI-3362b to a larger population of planets, plotting the sky-projected obliquity versus eccentricity for all giant planets with $e>0.1$. In the left panel, for single-star systems, we can see an emerging trend: sky-projected obliquities tend to be lower with increasing eccentricity. Systems with $0.1\lesssim e \lesssim 0.4$ are spreading from well-aligned to moderately misaligned ($\lambda \sim 45$ deg). All six systems with $e \gtrsim 0.4$ are well aligned, with TOI-3362b being the most eccentric one and having $\lambda<10$ deg at a $3\sigma$ confidence level thanks to ESPRESSO precision. This trend disappears when we consider binary stars, as shown in the right panel of Figure \ref{fig:scatter}. 

\subsection{Limits on potential outer companions}\label{subsec:companions}

The baseline of the RV measurements used in this work is $\sim2$ years, and residuals show no clear evidence of companions (see Figure \ref{fig:RV}). To estimate what kind of planets can produce RV signals consistent with the residuals of the RV model for TOI-3362b, we performed a population synthesis study.

We considered cold Jupiters with masses between 1 and 30 $M_J$, and with semimajor axes between 1 and 15 au. The eccentricity of the planets was drawn from a uniform distribution between 0.2 and 0.5, considering that cold Jupiters tend to be moderately eccentric \citep[e.g.,][]{Kipping2013}. The orbital inclinations of the companions were between 0 and 90 deg. With those distributions, we estimated the RV signals of different companions. Figure \ref{fig:companions} shows regions of the mass versus semimajor axis diagram where planets produced RV signals that are inconsistent with the residuals of Figure \ref{fig:RV}. We discard the existence of similar mass companions to TOI-3362b within $\sim 4.5$ au ($2\sigma$) and $\sim 7$ au ($1\sigma$). 

We also estimated the region of the diagram in Figure \ref{fig:companions} where secular eccentricity excitation is forbidden by relativistic precession. This occurs when the secular timescale due to the companion is longer than the time scale for relativistic precession. The ratio $\epsilon_{\rm gr}$ between these timescales can be calculated as:

\begin{equation}\label{eq:epsilon}
    \epsilon_{\rm gr} = \frac{G M_{\star}^2 a_c^3 (1-e_c^2)^{3/2}}{c^2 a^4 M_c},
\end{equation}
where $a_c$, $e_c$, and $M_c$ are the semimajor axis, eccentricity, and mass of the companion respectively. In Figure \ref{fig:companions}, the red dashed line corresponds to $\epsilon_{\rm gr} = 1$ for $e_c = 0.5$. 

\subsection{Origin}\label{subsec:origin}
The orbital properties of TOI-3362b and the limits for the presence of an outer companion pose a major challenge to most dynamical paths as we describe next.
\begin{itemize}
\item {\it In-situ formation or disk-driven migration} could naturally explain the small sky-projected obliquity, but does not naturally account for the high eccentricity of the orbit. The mass ratio $M_p/M_\star\simeq 2.5\times 10^{-3}$, may not be large enough to carve a wide and deep enough gap to drive significant eccentricity growth \citep{papa2001,Bitsch2013}. If, in turn, the planet resides inside a cavity, simulations show that the eccentricity may grow up to $\sim 0.4$ \citep{debras2021}, still well below the planet's eccentricity of 0.72.

\item {\it Planet-planet interactions} at the planet's current orbital location may have excited the planet's eccentricity either by scattering or secular excitation (e.g., \citealt{Anderson2017}) from a yet undetected companion. Scattering could account for the planetary eccentricity by ejecting planets (e.g., \citealt{anderson2020}), though such outcome is not expected to be dominant because the ratio between the planets' escape velocity and its circular velocity a the current semi-major axis is $\sim 1$ \citep{petro2014}. An added complication is to retain the small obliquity after scattering. In turn, the secular excitation is largely disfavored by our radial velocity measurements. As shown in Figure \ref{fig:companions}, our measurements exclude more massive companions inside $\sim 4-5$ au (at 2$\sigma$), while the potential planet allowed at larger orbital separations would be unable to drive secular excitation as relativistic precession would quench it (red dashed line). 

\item {\it Tidal high-eccentricity migration} can naturally account for the planetary eccentricity and semi-major axis. Assuming an equilibrium tide model, the planet would rapidly migrate from $\sim 1$ au (the circularization timescale from $e$ = 0.9 to 0.72 is $\sim0.15-1.5$ Gyr) and would stall as the eccentricity decreases (the circularization timescale from $e=$ 0.72 to 0.5 is $\sim 0.3-3$ Gyr). The small obliquity disfavors most high-eccentricity migration paths including von Zeipel-Lidov-Kozai oscillations \citep{Fabrycky07}, planet-planet scattering \citep{beauge2012}, and secular chaos \citep{Wu11}, while favoring CHEM \citep{Petrovich15} that can naturally account for the small obliquity.
\end{itemize}
In summary, a possible explanation for the system is CHEM driven by a yet undetected massive and distant companion. If found using longer-term observations, this model would predict that the companion should have a small to moderate mutual inclination ($ \lesssim 20$ deg), potentially measurable using upcoming Gaia astrometric data. Using the method\footnote{The method uses the Fisher matrix method to propagate the expected error of Gaia's astrometric and RV measurements into the parameters of different kinds of possible companions, thus constraining what signals can be detected. The relatively low probability of $\sim 30\%$ comes mostly from the fact that the star is massive and lies at $d\approx360$ pc, reducing the amplitude of the expected signal as observed by Gaia.}  from \citet{Espinoza-Retamal23} we estimate a probability of $\sim 30\%$ to measure the mutual inclinations if the planet lies at 5 au and use the current RV data. To improve our chances of detecting the potential planet and measuring the mutual inclination through a combined joint RV and Gaia astrometry analysis, long-term radial velocity monitoring will be necessary.

\begin{acknowledgements}
We would like to thank Jiayin Dong for useful discussions on the implementation of the RM effect in \texttt{exoplanet}.
JIER acknowledges support from the National Agency for Research and Development (ANID) Doctorado Nacional grant 2021-21212378. 
AJ, RB, CP, and MH acknowledge support from ANID -- Millennium  Science  Initiative -- ICN12\_009.
CP acknowledges support from CATA-Basal AFB-170002, ANID BASAL project FB210003, FONDECYT Regular grant 1210425, CASSACA grant CCJRF2105, and ANID+REC Convocatoria Nacional subvencion a la instalacion en la Academia convocatoria 2020 PAI77200076.
RB acknowledges support from FONDECYT project 11200751.
AJ acknowledges support from FONDECYT project 1210718.
GS acknowledges support provided by NASA through the NASA Hubble Fellowship grant HST-HF2-51519.001-A awarded by the Space Telescope Science Institute, which is operated by the Association of Universities for Research in Astronomy, Inc., for NASA, under contract NAS5-26555.
\end{acknowledgements}

\bibliography{sample631}{}
\bibliographystyle{aasjournal}

\appendix

\section{Corner plot}\label{ap:corner}

\begin{figure*}[h!]
    \centering
    \includegraphics[width = 18cm]{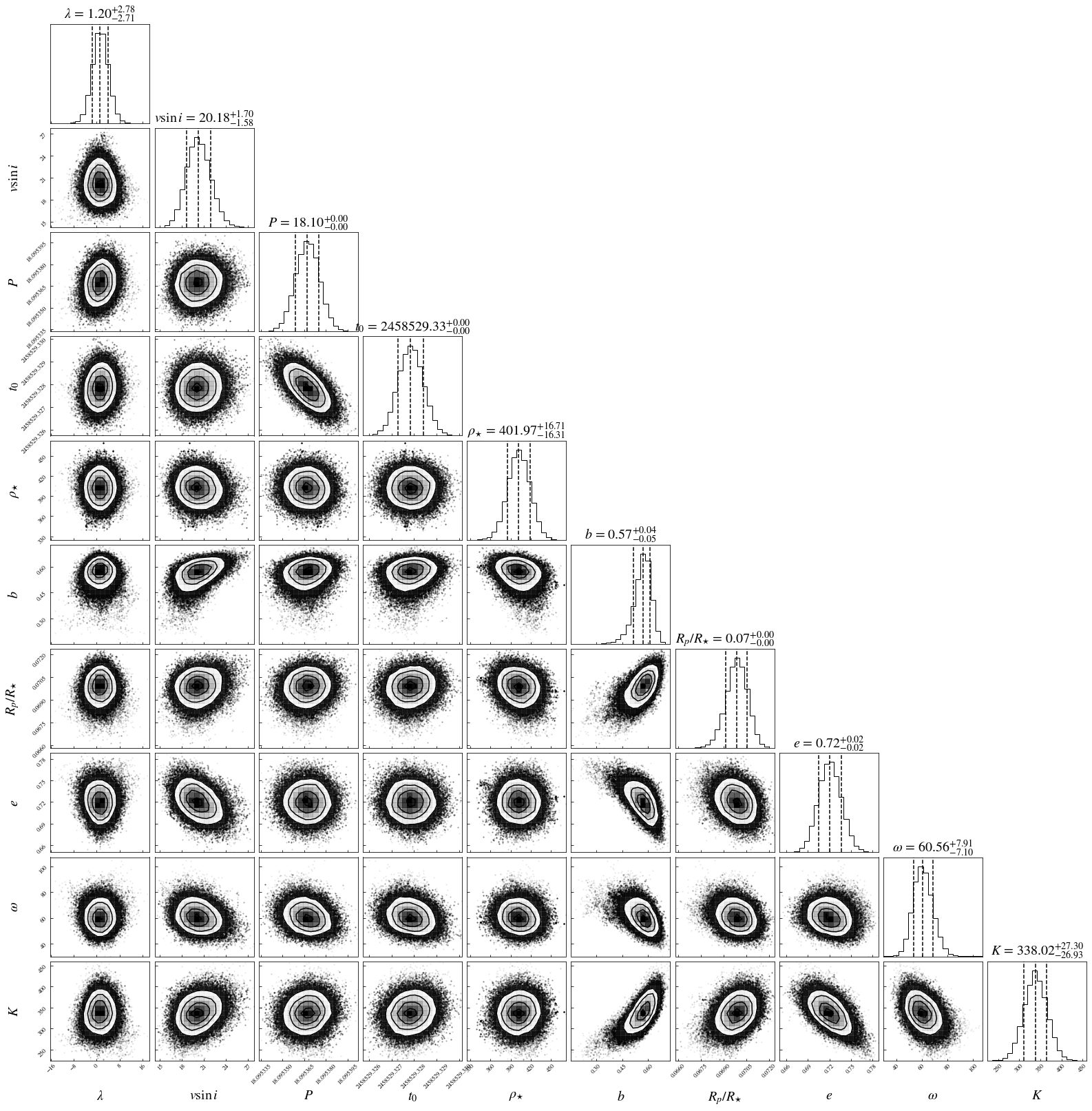}
    \caption{Joint posterior distributions and histograms of the posteriors for the physical parameters of our model.}
    \label{fig:corner}
\end{figure*}

\end{document}